\documentclass[11pt]{article}

\usepackage[a4paper,margin=1in]{geometry}

\usepackage[utf8]{inputenc}
\usepackage[T1]{fontenc}
\usepackage{xcolor}
\definecolor{docnotelinkcolor}{rgb}{0,0,0}

\usepackage{amsmath}
\usepackage{amssymb}

\usepackage[version=4]{mhchem}

\usepackage{float}
\usepackage{booktabs}
\usepackage{multirow}
\usepackage{tabularx}
\usepackage{graphicx}
\usepackage{subcaption}
\usepackage{siunitx}

\usepackage{algorithm}
\usepackage{algorithmicx}
\usepackage{algpseudocode}

\usepackage[hidelinks]{hyperref}
\usepackage{url}

\usepackage[numbers,sort&compress,super]{natbib}
\setcitestyle{super,open={},close={}}

\usepackage{caption}

\title{\textbf{Chemical filters for ultra-high-throughput materials screening and generation}}

\author{%
Kinga O. Mastej$^{1}$,
Panyalak Detrattanawichai$^{1}$,
Hyunsoo Park$^{1}$,
Anthony Onwuli$^{1}$,\\
Masahiro Negishi$^{1}$,
Aron Walsh$^{1,*}$\\[6pt]
\small $^{1}$Department of Materials, Imperial College London, London SW7 2AZ, United Kingdom\\
\small $^{*}$Corresponding author: a.walsh@imperial.ac.uk
}
\date{}

\begin{document}

\maketitle

\begin{abstract}
\noindent
Generative artificial intelligence is rapidly transforming materials design by enabling \textit{de novo} exploration of immense chemical spaces. Yet a large proportion of AI-generated compositions remain implausible, violating established chemical principles, which limits the reliability and interpretability of generative materials design. Here, we introduce a chemical validity operator that recasts heuristic chemical rules as a configurable algorithmic prior for evaluating and guiding generative materials discovery. Built on the open-source \texttt{SMACT} package, a data-informed oxidation-state model exposes tunable thresholds, allowing users to interpolate continuously between permissive and conservative chemical constraints, while supporting both exploratory and conservative materials-design workflows. Benchmarking six state-of-the-art generative models for inorganic crystals shows that most reproduce stoichiometry but under-represent realistic oxidation-state combinations, and that filtering removes compositions reliant on rarely observed oxidation states while preserving low-energy compounds near the convex hull. Beyond screening, the same operator can also serve as a reinforcement-learning reward, steering a latent diffusion model towards chemically grounded compositions. By encoding chemical heuristics and observations, this work establishes a foundation for oxidation-state-aware generative models.
\end{abstract}

\section{Introduction}

Recent advances in generative artificial intelligence (AI) have opened a new paradigm in computational materials design, enabling \textit{de novo} exploration of vast chemical spaces beyond existing databases \cite{zeni_generative_2025, handoko_artificial_2025}. Deep generative models such as variational autoencoders, diffusion models, and autoregressive models can now rapidly propose millions of candidate chemical compositions or crystal structures\cite{merchantScalingDeepLearning2023}, promising a substantial acceleration in hypothesis generation. A large proportion of generated candidates, however, violate fundamental principles of solid-state chemistry \cite{betalaLeMatGenBenchUnifiedEvaluation2026}, including charge neutrality, oxidation-state feasibility, and the electronegativity rule. Uncontrolled exploration of unphysical regions of chemical space poses a barrier to deployment, eroding trust in AI-guided discovery and wasting downstream computational and experimental effort on invalid compounds. 

An effective validity filter must be rapid enough to keep pace with generation, interpretable enough to justify its decisions, and general enough to accommodate the arbitrary compositions that generative models produce. A further requirement is less commonly recognised. Charge neutrality and the electronegativity rule constitute necessary but coarse criteria that most generated compositions already satisfy, and the finer distinction between plausible and implausible candidates lies in how closely their oxidation states conform to commonly observed chemistry. This distinction has direct consequences for experimental validation, since compounds that rely on rarely observed oxidation states tend to be more difficult to synthesise, often requiring bespoke routes or conditions rather than established methods \cite{acharySynthesisMaterialsUnusual2021}. A failed synthesis is correspondingly difficult to interpret, as it may reflect either an implausible candidate or merely an inadequate experimental route. The value of such compounds nonetheless depends on the objective, since rare oxidation states may represent a deliberate target in one workflow and an obstacle in another. As chemical plausibility is therefore a matter of degree rather than a binary property, a filter that returns only acceptance or rejection is insufficient. A more useful filter should afford adjustable control over how strongly empirical chemistry constrains its output. Oxidation states provide a particularly attractive basis for such a prior because they encode chemically interpretable constraints while remaining inexpensive to evaluate, making them well-suited to the large-scale screening and optimisation required by modern generative models.

Data-driven surrogate models offer one route to post-generation filtering, but such models have been shown to degrade beyond their training distribution and to provide limited interpretability \cite{dengSystematicSofteningUniversal2025, sanockiGeneralizationLongRangeMachine2026}. Chemical heuristics, by contrast, provide an interpretable means of enforcing physical plausibility by encoding established atomic and bonding principles. Established approaches include Goldschmidt's tolerance factor \cite{goldschmidtGesetzeKrystallochemie1926a} for filtering compositions on the basis of atomic radii and Pettifor structural maps \cite{pettiforChemicalScaleCrystalstructure1984} that exploit chemical similarity. Such approaches are, however, restricted to narrow structural families and cannot be applied to arbitrary compositions. Goldschmidt's tolerance factor, for example, applies only to \ce{ABX3} compositions of the perovskite structure type. These limitations may be overcome by generalising heuristic chemistry into algorithmic filters that are independent of composition and structure type. An early example is the use of the octet rule for valence-electron counting to identify potentially stable multi-component semiconductors \cite{pamplin_systematic_1964}. More recently, chemical and physical constraints have been embedded within the generative process itself rather than applied after sampling. CrysVCD \cite{cheng_enhancing_2025} generates charge-balanced compositions by construction, pairing each element with an oxidation state drawn from a fixed set of common states, while PGCGM \cite{zhao_physics_2023} imposes interatomic-distance and symmetry constraints during training. Such built-in constraints improve validity, but each is intrinsic to its host model, cannot assess compositions generated by other models, and applies a single fixed level of strictness.

The Semiconducting Materials by Analogy and Chemical Theory (SMACT) codebase was developed to explore large crystal chemical spaces at low computational cost \cite{davies_computational_2016}. Beginning from a set of chemical species and stoichiometric constraints, it enumerates the possible combinations and labels them according to established chemical rules. The package has since been extended from this enumeration role into a composition-level validity operator \cite{daviesMaterialsDiscoveryChemical2018, daviesSMACTSemiconductingMaterials2019}, and this operator has been widely adopted as a standard test of the chemical plausibility of compositions produced by generative models \cite{xie2022crystal, parkGuidingGenerativeModels2025a, qiuMassiveDiscoveryCrystal2025, yanMGBMaterialGeneration2025, weiCrystalCompositionTransformer2024}. In its established form, however, the operator returns a binary verdict based on predetermined oxidation-state assumptions, and has been applied almost exclusively as a passive screen on generated outputs.

This work develops the operator in two complementary respects. First, it allows the strictness of the chemical constraint to be adjusted, so that a user can decide how closely generated compositions must follow commonly observed oxidation-state chemistry, from permissive exploration to conservative screening. Second, we demonstrate that the operator can be incorporated into the generative process rather than used solely as a passive screen. To this end, it is first employed as a post-hoc diagnostic across six state-of-the-art generative models for inorganic crystals \cite{joshi2025allatom, xie2022crystal, park2025exploration, jiao2023crystal, jiao2024space, zeni_generative_2025}, quantifying how far each departs from experimentally dominant oxidation chemistry. Having established its value as an evaluation framework, we then employ the same operator as a reinforcement-learning reward to fine-tune Chemeleon2\cite{parkGuidingGenerativeModels2025a}, a pretrained latent diffusion model, in both binary and continuous forms. Together, these developments establish a unified, oxidation-state-aware chemical-feasibility framework that not only evaluates generated compositions but also steers their generation.

\section{Results}

\subsection{A chemical validity operator for generative screening}

\begin{figure}[htbp]
    \centering
    \includegraphics[width=\linewidth]{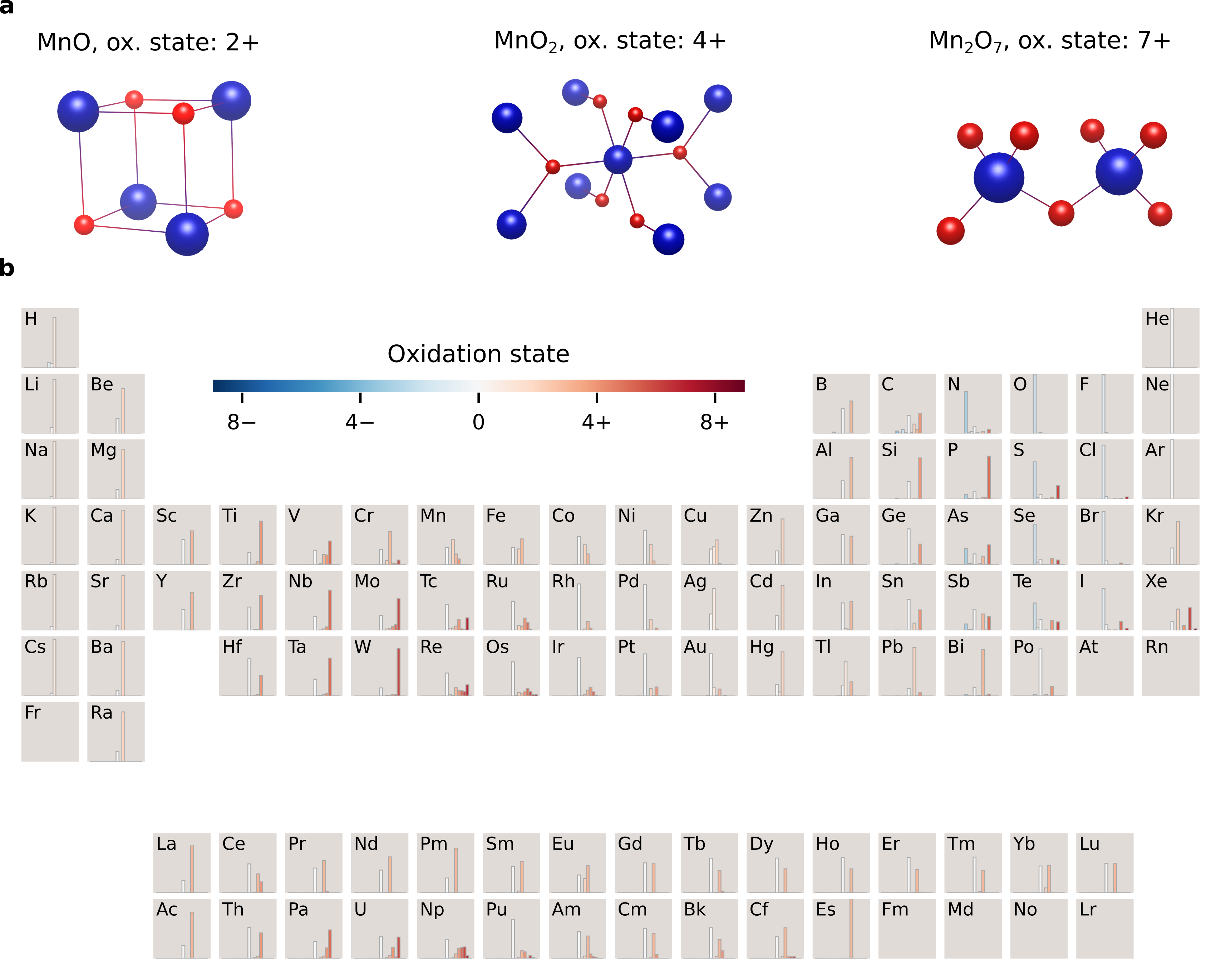}
    \caption{Oxidation state distributions across the periodic table. (a) Representative manganese oxides illustrating three accessible oxidation states: MnO ($\mathrm{Mn}^{2+}$), $\mathrm{MnO_2}$ ($\mathrm{Mn}^{4+}$), and $\mathrm{Mn_2O_7}$ ($\mathrm{Mn}^{7+}$). (b) Bar charts of the relative frequency of oxidation states for each element, derived from the Inorganic Crystal Structure Database \cite{zagorac_recent_2019}; colour encodes the oxidation state value (red: positive, blue: negative).}
    \label{fig1}
\end{figure}

Most generative models for inorganic materials operate at the level of a crystal structure, producing atom types together with their coordinates and the unit cell of the periodic lattice \cite{xie2022crystal, jiao2023crystal, jiao2024space, zeni_generative_2025, joshi2025allatom, park2025exploration}. In each case, the identity of every atom is generated as a bare element, with no explicit notion of oxidation state. Chemical feasibility, however, is frequently determined at the level of species. The same element can correspond to several chemically distinct electronic arrangements, and not all of these are compatible with a given formula. Oxidation states therefore provide a bridge between the element-level outputs of generative models and the valence constraints that govern whether a proposed composition is chemically plausible.

Manganese illustrates this distinction. The element Mn can appear as Mn$^{2+}$ in MnO, Mn$^{4+}$ in MnO$_2$, and Mn$^{7+}$ in Mn$_2$O$_7$. These species correspond to markedly different oxide chemistries (Figure~\ref{fig1}a), differing in electron count, bonding preference, coordination
environment and reactivity. The oxidation-state preferences are highly non-uniform across the elements of the periodic table (Figure~\ref{fig1}b). Many main-group elements are dominated by one or two oxidation states, whereas transition-metal and f-block elements often exhibit broader, multimodal distributions.

Importantly, the empirical distributions are not simply binary lists of allowed
and forbidden states. Common oxidation states form dominant peaks, while rarer
states appear as low-frequency, chemically-specific tails, which may require special structural stabilisation or external conditions such as temperature or pressure. This graded
pattern provides the basis for a tunable chemical-validity prior. Rather than
imposing a fixed boundary between valid and invalid chemistry, the prior governs the extent to which rare chemistry is retained when screening or steering
generated compositions.

\subsection{Chemical validity rates across six generative models}

The data-driven filters were applied to six generative models for inorganic materials: ADiT \cite{joshi2025allatom}, CDVAE \cite{xie2022crystal}, Chemeleon-DNG \cite{park2025exploration}, DiffCSP \cite{jiao2023crystal}, DiffCSP++ \cite{jiao2024space}, and MatterGen \cite{zeni_generative_2025}. For each model, 10,000 compositions were generated unconditionally, duplicates retained, and screened under increasingly strict criteria (Table~\ref{table_nonunique}); full screening settings are given in the Methods. Only charge neutrality ($q$) together with the electronegativity rule ($q+\chi$) act as necessary validity criteria. The consensus and commonality filters are not definitions of validity but increasingly conservative empirical priors, which quantify how reliably a given generative model reproduces the experimental oxidation-state distribution. Lower retention reflects greater sampling of rare, yet not forbidden, oxidation states, rather than a higher rate of chemically impossible compounds.

\begin{table}[ht]
    \centering
    \caption{SMACT chemical validity of 10{,}000 compositions sampled from each model without property conditioning; values are the percentage of compositions retained. Filters are cumulative, so each column applies every constraint to its left and the one in its header. $q$: a charge-neutral assignment exists using a feasible oxidation state for each element. $q+\chi$: that assignment also satisfies the Pauling electronegativity rule (Eq.~\ref{eq:pauling}). $C{=}3$: it uses only oxidation states observed in at least three ICSD entries (consensus, Eq.~\ref{eq:consensus}). The commonality columns require, in addition, that each oxidation state account for at least a fraction $\theta$ of all ICSD occurrences recorded for that element (Eq.~\ref{eq:commonality}), e.g.\ $\theta=0.5$ keeps only states making up at least half of an element's occurrences.}
    \label{table_nonunique}
    \begin{tabular}{lccccccc}
    \toprule
     & & & & \multicolumn{3}{c}{Commonality $\theta$} \\
    \cmidrule(lr){5-7}
    Model & $q$ & $q+\chi$ & $C{=}3$ & $0.05$ & $0.1$ & $0.5$ \\
    \midrule
    MP-20 baseline\cite{jain_commentary_2013, xie2022crystal} & 96.5\% & 94.6\% & 93.2\% & 78.1\% & 73.6\% & 59.7\% \\
    ADiT\cite{joshi2025allatom}               & 96.5\% & 94.8\% & \textbf{93.1\%} & \textbf{76.2\%} & \textbf{71.5\%} & \textbf{54.9\%} \\
    CDVAE\cite{xie2022crystal}                & \textbf{97.3\%} & \textbf{95.6\%} & 91.4\% & 59.3\% & 53.4\% & 41.8\% \\
    Chemeleon-DNG\cite{park2025exploration}   & 93.6\% & 91.4\% & 87.7\% & 61.4\% & 57.0\% & 47.9\% \\
    DiffCSP\cite{jiao2023crystal}             & 93.1\% & 90.8\% & 86.8\% & 60.2\% & 57.2\% & 48.9\% \\
    DiffCSP++\cite{jiao2024space}             & 95.8\% & 93.8\% & 90.0\% & 61.0\% & 56.5\% & 47.2\% \\
    MatterGen\cite{zeni_generative_2025}      & 95.0\% & 92.5\% & 88.6\% & 60.7\% & 56.4\% & 47.7\% \\
    \bottomrule
    \end{tabular}
\end{table}

As shown in Table~\ref{table_nonunique}, across all models 93--97\% of generated compositions pass charge neutrality and admit at least one oxidation-state assignment. When the electronegativity rule is applied ($q+\chi$), a further 2--5\% of compositions are rejected owing to chemically inconsistent redox assignments. Statistical oxidation-state filtering ($C=3$) imposes a stronger empirical prior, reducing the fraction of retained compositions by a further 3--7\%. The effect is strongest under the strictest commonality filter ($\theta=0.5$), where only 42--55\% of compositions are retained. The MP-20 dataset comprises Materials Project structures with no more than 20 atoms per unit cell and lies within the training distribution of all six models. It therefore provides a reference against which the oxidation-state statistics reproduced by each generator can be compared. As shown in Table~\ref{table_nonunique}, all models closely track this reference under the weakest constraints. Under the $q+\chi$ criterion, the retention fractions range from 90.8\% to 95.6\%, compared with 94.6\% for MP-20. This indicates that stoichiometric charge balance and qualitative redox ordering are broadly captured by the generated compositions. More substantial differences emerge only when the empirical oxidation-state prior is made more restrictive. At $\theta=0.5$, model retention falls to 41.8--54.9\%, compared with 59.7\% for MP-20. ADiT remains closest to the reference distribution (54.9\%), CDVAE shows the largest deviation (41.8\%), and the remaining four models cluster between 47\% and 49\%. Because the stricter filters remove statistically rare oxidation states rather than charge-forbidden compositions, retention under these criteria measures
oxidation-state commonality, not absolute chemical validity. It therefore
quantifies how closely each generator follows the dominant oxidation chemistry
of MP-20, as opposed to sampling rarer but documented environments. Retention
fractions were estimated from 10,000 generated compositions, giving a maximum
binomial standard error of 0.5\%.

\begin{table}[ht]
    \centering
    \caption{SMACT chemical validity for unique compositions only (7{,}000 samples); values are the percentage of compositions retained. The table demonstrates the robustness of the trends in Table~\ref{table_nonunique} to polymorph duplication. Columns and filters are defined as in Table~\ref{table_nonunique}.}
    \label{table_unique}
    \begin{tabular}{lccccccc}
    \toprule
     & & & & \multicolumn{3}{c}{Commonality $\theta$} \\
    \cmidrule(lr){5-7}
    Model & $q$ & $q+\chi$ & $C{=}3$ & $0.05$ & $0.1$ & $0.5$ \\
    \midrule
    MP-20 baseline\cite{jain_commentary_2013, xie2022crystal} & 96.5\% & 94.3\% & 92.6\% & 76.9\% & 72.0\% & 58.9\% \\
    ADiT\cite{joshi2025allatom}               & 96.5\% & 94.7\% & \textbf{92.9\%} & \textbf{76.1\%} & \textbf{71.2\%} & \textbf{55.8\%} \\
    CDVAE\cite{xie2022crystal}                & \textbf{97.2\%} & \textbf{95.4\%} & 91.1\% & 58.8\% & 52.9\% & 41.2\% \\
    Chemeleon-DNG\cite{park2025exploration}   & 93.6\% & 91.5\% & 87.7\% & 60.6\% & 56.1\% & 46.8\% \\
    DiffCSP\cite{jiao2023crystal}             & 93.1\% & 90.7\% & 86.3\% & 59.1\% & 56.1\% & 47.9\% \\
    DiffCSP++\cite{jiao2024space}             & 95.5\% & 93.5\% & 89.6\% & 60.0\% & 55.3\% & 46.1\% \\
    MatterGen\cite{zeni_generative_2025}      & 94.9\% & 92.3\% & 88.4\% & 60.3\% & 55.8\% & 46.8\% \\
    \bottomrule
    \end{tabular}
\end{table}

\begin{figure}[htbp]
    \centering
    \includegraphics[width=\textwidth]{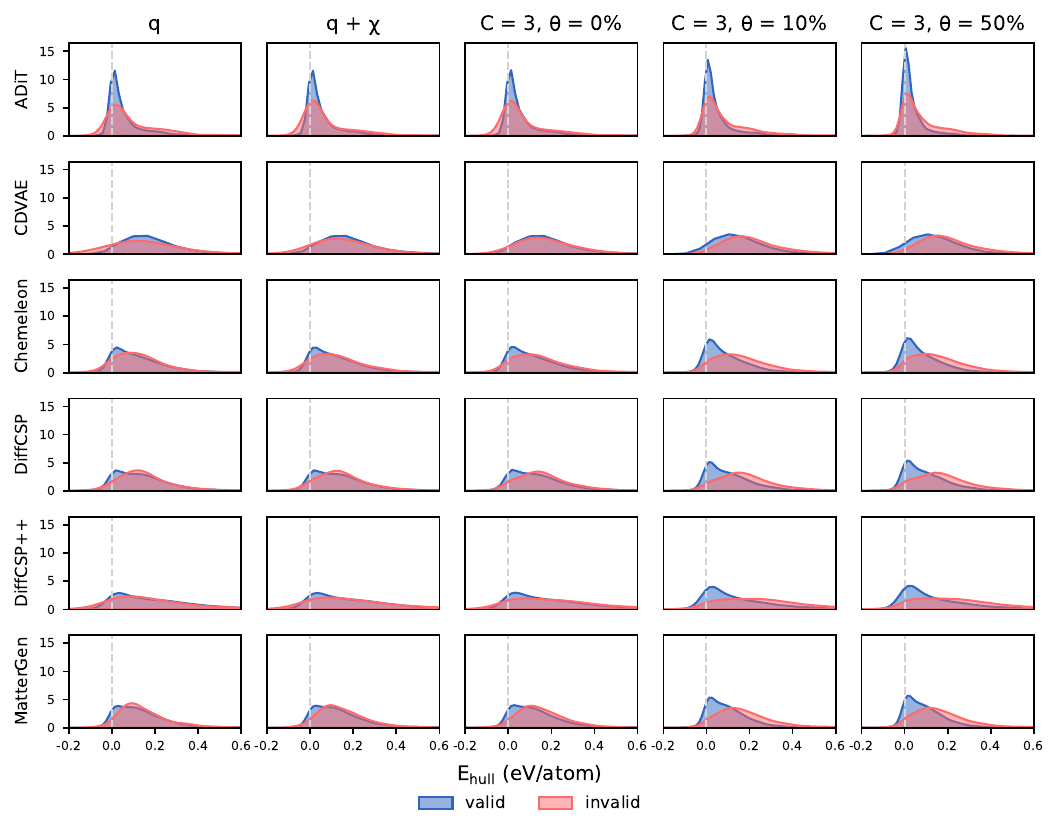}
    \caption{Distributions of energy above the convex hull ($E_{hull}$) for compositions generated by six generative models under progressively stricter \texttt{SMACT} validity definitions. For each model, generated compositions are stratified \textit{post hoc} into valid (blue) and invalid (pink) subsets under the specified chemical criterion, without altering the underlying generative distribution. While the overall shape and central tendency of the $E_{hull}$ distributions remain largely unchanged across validity definitions, \texttt{SMACT}-invalid compositions are increasingly enriched in the high-energy tail as more conservative oxidation-state priors are applied. By contrast, valid compositions consistently populate the low-$E_{hull}$ region near the convex hull across all filtering levels, which indicates a correlation between chemical inconsistency and thermodynamic instability rather than an explicit stability bias imposed by the strict chemical filters.}
    \label{ehulls}
\end{figure}

The same qualitative trends are observed when duplicate structures corresponding to identical compositions (i.e.\ polymorphs) are removed (Table~\ref{table_unique}), confirming that the conclusions reflect compositional rather than structural sampling biases. The relationship between chemical plausibility and thermodynamic accessibility is examined next (Figure~\ref{ehulls}), quantified by the energy above the convex hull ($E_{hull}$) defined as the energy of a composition relative to the most stable combination of competing phases, where $E_{hull}=0$ marks a thermodynamically stable compound on the hull and larger values indicate progressively less stable, metastable or unstable compounds. Stratifying the generated compositions by \texttt{SMACT} validity reveals a systematic correlation between oxidation-state inconsistency and thermodynamic instability. Across all models, \texttt{SMACT}-valid compositions occupy similar regions of the $E_{hull}$ distribution irrespective of the validity definition applied, while \texttt{SMACT}-invalid compositions are increasingly concentrated in the high-$E_{hull}$ tail under more conservative oxidation-state priors. Quantitatively, this separation grows with filter strictness. Under the charge neutrality filter alone ($q$) the median $E_{hull}$ of invalid and valid compositions differs by only $0.011$~eV/atom, whereas at the strictest setting ($C=3$, $\theta=0.5$) the median $E_{hull}$ of SMACT-invalid compositions ($0.152$~eV/atom) exceeds that of valid compositions ($0.063$~eV/atom) by $0.089$~eV/atom, with $65\%$ of invalid compositions lying above $0.1$~eV/atom\cite{sunThermodynamicScaleInorganic2016} compared with $38\%$ of valid ones (Mann--Whitney $U$ test, $p < 10^{-300}$, $n = 58{,}469$). Importantly, this behaviour reflects correlations within the generative outputs rather than any effect of the screening procedure itself, since \texttt{SMACT} is applied strictly as a \textit{post hoc} diagnostic and does not modify the sampling process.

A single-core timing benchmark using the same 10{,}000-composition sets and filtering path as Table~\ref{table_nonunique} required 121 to 262 seconds per model and filter setting, corresponding to 38 to 82 compositions per second. These timings are conservative because the current implementation reconstructs the oxidation-state table on each call. The filters are therefore fast enough for routine post hoc evaluation of benchmark-scale generated composition sets, with full timing details reported in Supplementary Table~\ref{tab:screening_cost}.

\subsection{Tunable chemical control}

\begin{figure}[!htbp]
    \centering
    \includegraphics[width=\linewidth]{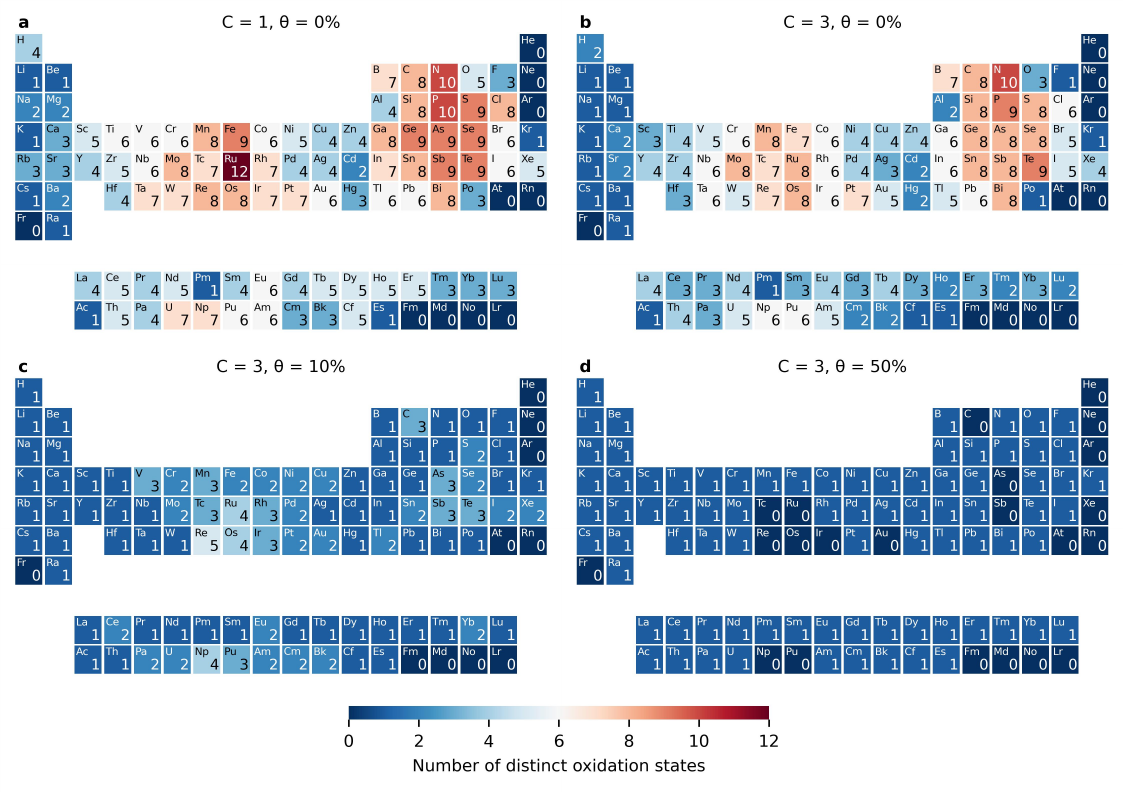}
    \caption{Number of oxidation states retained for each element under four filtering settings. The permissive setting retains 490 element oxidation-state pairs across 93 elements. Increasing the consensus threshold from \(C=1\) to \(C=3\) removes weakly supported states, while increasing the commonality threshold \(\theta\) suppresses states that are rare relative to the other recorded states of the same element. At \(C=3\) and \(\theta=0.50\), 81 element oxidation-state pairs remain and each retained element has a single accessible oxidation state. The colour and number in each tile indicate how many oxidation states remain for that element, with navy tiles marking elements for which no oxidation state survives.}
    \label{fig:element_frequency1}
\end{figure}

Figure~\ref{fig:element_frequency1} shows how the filtering thresholds alter the oxidation-state search space before complete compositions are evaluated. Under the most permissive setting, the ICSD-derived table contains 490 element oxidation-state pairs across 93 elements. Increasing the consensus threshold from \(C=1\) to \(C=3\) at \(\theta=0\) removes 85 pairs, corresponding to a 17.3\% reduction, while retaining the same number of elements with at least one accessible state. The larger contraction occurs when the commonality threshold is introduced. Raising \(\theta\) to 0.10 removes a further 256 pairs, and the strictest setting retains only 81 pairs across 81 elements. At this setting, every retained element has exactly one accessible oxidation state. The reduction is therefore not uniform across the periodic table. The \(p\) and \(d\) blocks contain the largest number of retained pairs under permissive filtering, with 181 and 178 pairs respectively, but collapse to 22 and 24 pairs under the strictest setting. This establishes that the commonality threshold acts primarily by suppressing rare alternative oxidation states for multivalent elements, rather than by applying a uniform reduction to all elements.

\begin{figure}[!htbp]
    \centering
    \includegraphics[width=\linewidth]{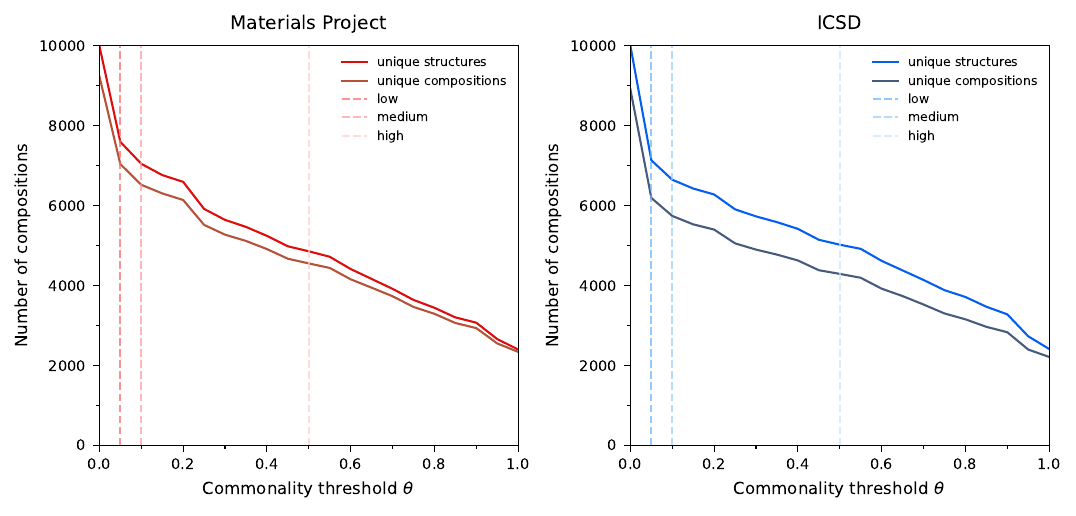}
    \caption{Retention of Materials Project and ICSD entries as a function of the oxidation state commonality threshold. Solid lines show counts for unique structures and unique reduced compositions. Vertical dashed lines mark the low, medium, and high threshold settings at \(\theta=0.05\), \(\theta=0.10\), and \(\theta=0.50\). A threshold of \(\theta=0\) corresponds to the unfiltered reference set, while increasing \(\theta\) requires each oxidation state in an accepted composition to account for a larger fraction of the ICSD occurrences recorded for that element.}
    \label{fig:filter_retention}
\end{figure}

Figure~\ref{fig:filter_retention} shows how the species level pruning in Figure~\ref{fig:element_frequency1} propagates to complete compositions in reference databases. Both show a smooth decrease in retention as \(\theta\) increases, indicating that the commonality threshold provides graded control over the admitted composition space. The largest loss occurs immediately when the threshold is raised from \(\theta=0\) to \(\theta=0.05\), which removes 2402 Materials Project structures and 2863 ICSD structures. This corresponds to losses of 24.0\% and 28.6\%, respectively, and shows that a small commonality requirement removes many compositions containing very rare oxidation states. Further increases in \(\theta\) continue to contract the retained space more gradually. At the high setting of \(\theta=0.50\), 4859 Materials Project structures and 5024 ICSD structures are retained, corresponding to 48.6\% and 50.2\% of the initial structure sets. The unique structure and unique composition curves follow the same trend, with the largest separation already present at \(\theta=0\), indicating that the threshold response is driven primarily by oxidation state chemistry rather than polymorph duplication.

\begin{figure}[!htbp]
    \centering
    \includegraphics[width=\linewidth]{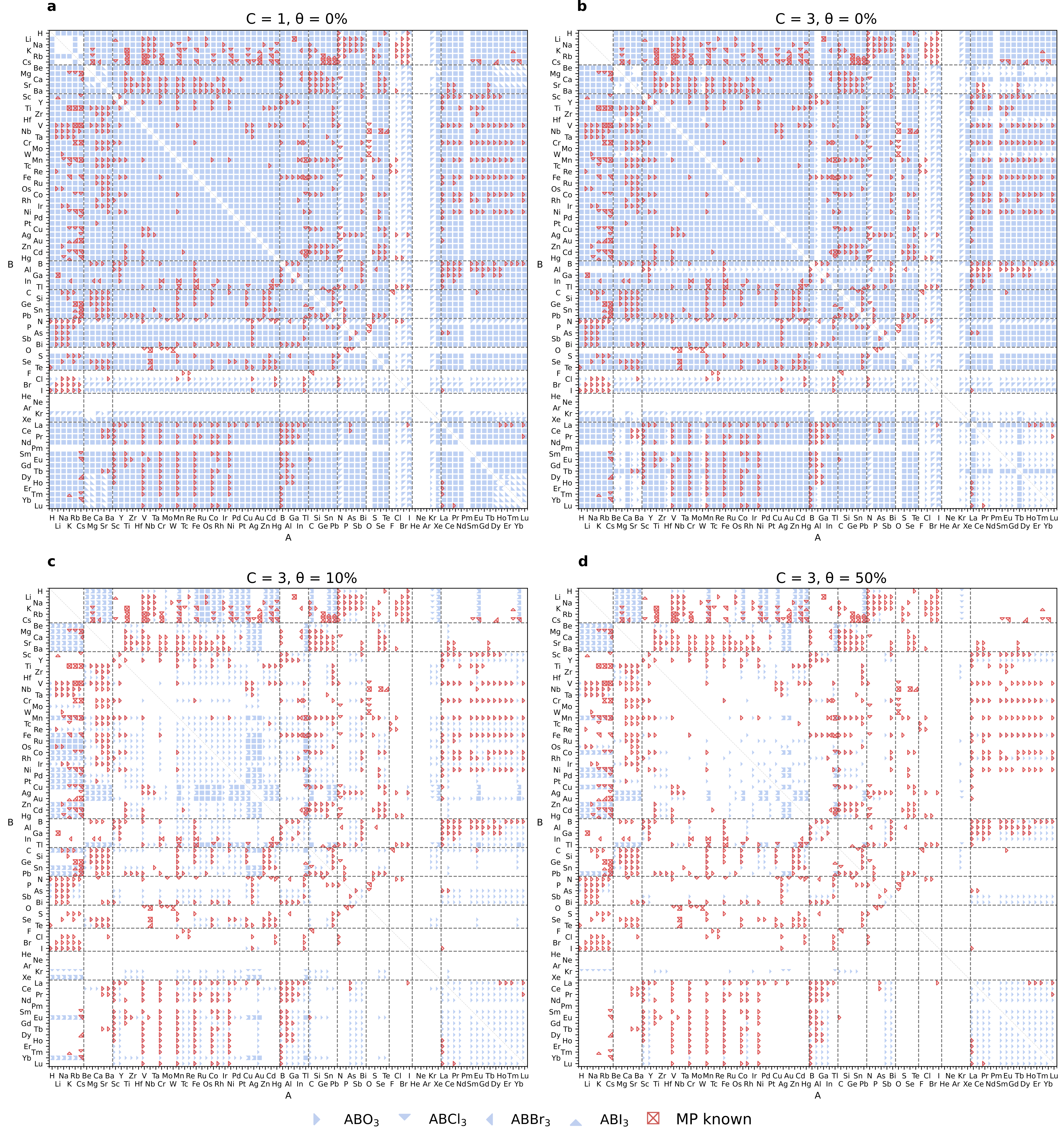}
    \caption{Pruning of the \(\mathrm{ABX_3}\) perovskite compositional design space under progressively stricter \texttt{SMACT} filtering settings. Each panel shows admissible A and B element combinations for oxide and halide perovskites with \(X=\mathrm{O}\), \(\mathrm{Cl}\), \(\mathrm{Br}\), or \(\mathrm{I}\). Blue markers denote compositions satisfying the specified \texttt{SMACT} criteria, while red markers indicate experimentally reported perovskites from the Materials Project. Across the four settings, the number of admissible compositions decreases from 11091 to 781, while the fraction of admissible compositions corresponding to known Materials Project perovskites increases from 5.4\% to 29.4\%.}
    \label{fig:four_panel_figure}
\end{figure}

Figure~\ref{fig:four_panel_figure} tests how the same filtering choices act in a chemically interpretable \(\mathrm{ABX_3}\) design space. Under the most permissive setting, \texttt{SMACT} admits 11091 oxide and halide compositions. Raising the consensus threshold from \(C=1\) to \(C=3\) removes 787 compositions, whereas introducing \(\theta=0.10\) removes a further 8371 compositions. The main contraction therefore arises from suppressing oxidation states that are rare relative to the dominant states of the same element, rather than from removing states with limited absolute database support alone. At the strictest setting, 781 compositions remain, corresponding to a 93.0\% reduction relative to the permissive space. Experimentally reported Materials Project perovskites are also removed by stricter filtering, with their retention decreasing from \(90.7\%\) to \(34.9\%\). However, the broader hypothetical design space contracts more strongly, so known perovskites constitute a larger fraction of the retained set, increasing from \(5.4\%\) to \(29.4\%\) of possible compositions. The filter therefore does not merely reduce the number of candidates. It preferentially contracts the search space towards compositions supported by common formal valence assignments, while retaining a smaller set of known perovskites and additional candidates for downstream structural and thermodynamic screening.

Together, these analyses show that the consensus and commonality thresholds provide graded and chemically structured control over the admitted chemical space. The largest contraction arises from commonality filtering, which suppresses rare oxidation states for multivalent elements and propagates from the species table to reference databases and the \(\mathrm{ABX_3}\) design space. The retained space is therefore not reduced uniformly, but enriched in compositions supported by common formal valence assignments.

\subsection{Reinforcement learning with chemically informed rewards}

\begin{figure}[htbp]
    \centering
    \includegraphics[width=\linewidth]{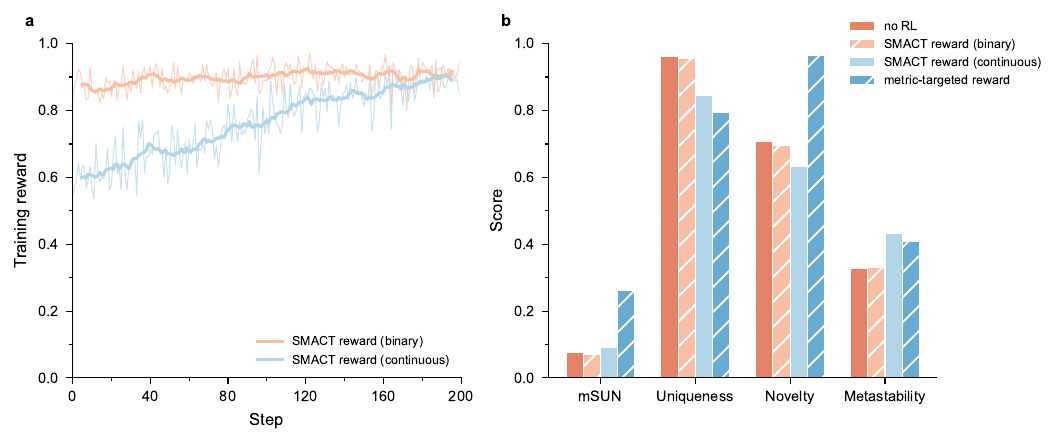}
    \caption{Reinforcement learning of \texttt{Chemeleon2} with chemically informed rewards. (a) Training reward trajectories for binary and continuous \texttt{SMACT} rewards. The binary reward saturates rapidly, whereas the continuous reward increases more gradually because it retains information about oxidation state commonality. (b) Downstream performance of the pretrained model, the two \texttt{SMACT} reward models, and the evaluation metric reward model across mSUN, uniqueness, novelty, metastability. The continuous \texttt{SMACT} reward increases mSUN and metastability relative to the pretrained model, while the evaluation metric reward gives the highest mSUN and novelty.}
    \label{fig:rl_reward}
\end{figure}

Figure~\ref{fig:rl_reward}(a) shows that the binary and continuous rewards provide different optimisation signals. The binary reward is already high before fine tuning, increasing only from 0.834 to 0.940 over the plotted interval and reaching a plateau at step 56. This indicates that most sampled compositions already satisfy the binary \texttt{SMACT} validity criterion, so the reward rapidly loses resolution as a training objective. By contrast, the continuous reward increases from 0.538 to 0.849 and reaches a plateau only near the end of training, with a final 20 step mean of 0.894. The continuous formulation therefore provides a more informative optimisation signal because it continues to distinguish between valid compositions according to oxidation state commonality.

Figure~\ref{fig:rl_reward}(b) shows the downstream effect of these objectives. The binary reward does not improve the overall generative metrics relative to the pretrained model, with mSUN, the fraction of generated structures that are simultaneously metastable, unique, and novel (Methods), decreasing from 0.0764 to 0.0703, uniqueness from 0.9619 to 0.9544, and novelty from 0.7082 to 0.6958. Metastability changes only marginally, from 0.3279 to 0.3298. This is consistent with the saturation observed during training, since an already high binary validity score provides little additional information for reshaping the generated distribution. The continuous reward gives a clearer effect, increasing mSUN to 0.0908 and metastability to 0.4310, while reducing uniqueness to 0.8446 and novelty to 0.6315. Thus, the graded oxidation-state reward shifts the model towards more metastable outputs, but at the cost of sample diversity and novelty.

The evaluation metric reward gives the largest mSUN and novelty values, reaching 0.2604 and 0.9650, respectively, because the training objective is directly aligned with the reported evaluation metrics. However, this improvement is accompanied by lower uniqueness than the pretrained model, at 0.7947, and lower metastability than the continuous \texttt{SMACT} reward, at 0.4090 compared with 0.4310. These results show that \texttt{SMACT}-based rewards do not simply reproduce metric-targeted optimisation. The binary reward is limited by saturation, whereas the continuous reward provides a chemically defined training signal that can alter the balance between metastability, uniqueness, and novelty.

\section{Discussion}

This work develops \texttt{SMACT} from a simple composition enumerator into a tunable screening tool for generative materials design. By combining charge neutrality, oxidation-state feasibility, electronegativity ordering, and empirical oxidation-state statistics, the framework allows us to evaluate the outputs of generative models as well as steer generation towards oxidation states that are frequently observed in experimental crystal-structure data. This distinction is important, as a candidate may be charge-balanced and chemically possible while still relying on rare oxidation states that are weakly represented in available experimental data. In this form, \texttt{SMACT} provides a fast, interpretable, and model-agnostic layer between unconstrained generative sampling and more expensive thermodynamic or structural evaluation.

Benchmarking across six generative crystal models shows that charge-balanced stoichiometry is largely captured by contemporary architectures, whereas agreement with empirical oxidation-state commonality varies more substantially. The stricter consensus and commonality filters therefore quantify how closely a generated distribution follows dominant oxidation-state chemistry in experimental reference data. This makes \texttt{SMACT} useful as a diagnostic supplement to conventional generative metrics such as uniqueness, novelty, and metastability, which measure distributional or energetic outcomes but do not directly test whether a formula is chemically plausible at the level of formal valence assignments. The adjustable consensus and commonality thresholds provide a mechanism for navigating the realism--novelty trade-off. Permissive settings retain rare documented oxidation environments and are appropriate when the aim is broad exploration, including candidates that may require unusual stabilisation mechanisms or synthesis conditions. Stricter settings favour compositions built from oxidation states with stronger empirical support and are better suited to conservative screening or synthesis-oriented downselection. The perovskite design-space analysis illustrates that this control is chemically justified as filtering preferentially removes charge-inconsistent regions of the configurational space while retaining known families and plausible unexplored compositions. Taken together, the \(E_{\mathrm{hull}}\) results support this interpretation, with \texttt{SMACT}-invalid compositions substantially more likely than valid compositions to lie above the \(0.1\) eV atom\(^{-1}\) metastability threshold under the strictest filter, at \(65\%\) compared with \(38\%\).

The reinforcement learning results show how the \texttt{SMACT} formalism can be used to construct chemical objectives with different levels of resolution. The binary \texttt{SMACT} reward reaches a high training value rapidly, consistent with a pass or fail validity test that provides a strong and direct optimisation signal. Once a generated composition satisfies the validity operator, however, this reward does not distinguish between common and rare oxidation state assignments. The continuous reward addresses this limitation by ranking charge neutral assignments according to the experimental frequency of the oxidation states involved. This additional resolution gives the model a richer chemical signal, improving mSUN and metastability at the cost of reduced uniqueness and novelty. This trade-off is expected as steering generation towards commonly observed oxidation chemistry necessarily concentrates sampling in better-charted regions of composition space, mirroring the realism–novelty trade-off that the tunable thresholds expose in the screening setting. The evaluation metric reward achieves the highest novelty, as expected from its direct alignment with the reported metrics, but the \texttt{SMACT} rewards offer a more interpretable route for steering generation. These results suggest that oxidation state based rewards are most useful as chemically grounded auxiliary objectives that can complement direct optimisation of stability, diversity, or novelty. Because \texttt{SMACT} rewards are independent of the underlying generator, they can serve either as a post hoc filter on sampled outputs, as in the screening experiments, or as a model-agnostic in-training reward, as is the case here. This portability lets oxidation-state-aware objectives be layered onto current and future generative architectures without retraining or architectural modification.

Nonetheless, the framework remains bounded by the information it encodes. It operates at the composition level and therefore does not evaluate coordination geometry, site ordering, local bonding motifs, magnetism, pressure- or temperature-stabilised phases, or kinetic accessibility. Its empirical priors also inherit the coverage and reporting biases of the underlying crystallographic data, and stringent thresholds can penalise rare but real chemistry. These limitations define the appropriate role of the method as \texttt{SMACT} is a lightweight chemical prior, not a complete model of materials formation. Coupling oxidation-state-aware filtering with structure-level validity tests, learned thermodynamic models, synthesisability predictors, and electronic-structure feedback offers a route towards generative materials workflows that are both scalable and more faithful to the chemical constraints governing inorganic compounds.


\section{Methods}

\subsection{Oxidation states set}

The allowed set $\Omega_{\text{filtered}}$ is built from Inorganic Crystal Structure Database (ICSD) oxidation-state frequencies using two filters, where each candidate state is a species $E^q$ ($E$ the element, $q$ its oxidation state). The consensus filter removes states that appear infrequently in the ICSD,
\begin{equation}
    \delta_C(E^q) =
    \begin{cases}
        1 & \text{if } f(E^q) \ge C, \\
        0 & \text{otherwise,}
    \end{cases}
    \label{eq:consensus}
\end{equation}
where $f(E^q)$ is the number of ICSD occurrences of $E^q$ and $C$ is the consensus threshold. For example, the default $C=3$ removes species seen only once or twice while retaining those with support from multiple studies. The commonality filter removes states that are rare relative to other states of the same element,
\begin{equation}
    \gamma_\theta(E^q) =
    \begin{cases}
        1 & \textstyle\text{if } \frac{f(E^q)}{\sum_{q'} f(E^{q'})} \ge \theta, \\
        0 & \text{otherwise,}
    \end{cases}
    \label{eq:commonality}
\end{equation}
comparing each species against the total occurrences of its element rather than an absolute count. For example, $\theta = 0.1$ retains a state only if it accounts for at least one tenth of that element's recorded occurrences, while increasing $\theta$ tightens the prior towards each element's dominant oxidation chemistry. The filtered set is 
\begin{equation}
    \Omega_{\text{filtered}} = \{ E^q \; | \; \delta_C(E^q) = 1 \wedge \gamma_\theta(E^q) = 1 \} .
\end{equation}

\subsection{Chemical validity screening}

In the \texttt{SMACT} framework a chemical formula is first reduced to its stoichiometric form $c=\sum_i x_i E_i$, where $x_i$ correspond to the stoichiometric coefficients and $E_i$ to the distinct elements. Subsequently, \texttt{SMACT} enumerates every candidate assignment $(q_1,\dots,q_n)$ of elemental charge drawn from the allowed set of oxidation states $\Omega_{\text{filtered}}(E_i)$,
\begin{equation}
    q_i \in \Omega_{\text{filtered}}(E_i) \quad \forall i .
    \label{eq:qi_in_omega}
\end{equation}
A valid combination must meet further requirements of charge neutrality,
\begin{equation}
    \sum_i x_i q_i = 0 .
    \label{eq:charge_neutrality}
\end{equation}
and the electronegativity rule,
\begin{equation}
    (\chi_i - \chi_j)(q_i - q_j) \le 0 \quad \forall i,j ,
    \label{eq:pauling}
\end{equation}
where $\chi_i$ is the Pauling electronegativity of element $i$. 
A composition is valid only when all three constraints are satisfied,
\begin{equation}
    \mathbf{V}(c) =
    \begin{cases}
        1 & \text{if Eqs.~\eqref{eq:qi_in_omega}--\eqref{eq:pauling} are satisfied,} \\
        0 & \text{otherwise.}
    \end{cases}
\end{equation}
For example, $\mathrm{Mg_2O}$ is rejected because none of the candidate assignments is
charge-neutral at the $2\!:\!1$ ratio. The only allowed states are $\mathrm{Mg}^{2+}$ and
$\mathrm{O}^{2-}$, and the states that would balance the formula ($\mathrm{Mg}^{+}$ or
$\mathrm{O}^{4-}$) are electronically unstable and hence absent from $\Omega_{\text{filtered}}$.

Full pseudocode for the chemical validity screening is given in Algorithm~\ref{alg:smact_validity}.

\begin{algorithm}[H]
\caption{\texttt{SMACT} chemical validity screening algorithm}
\label{alg:smact_validity}
\begin{algorithmic}[1]
\State \textbf{Input:} composition $c$
\State \textbf{Parameters:} \textit{use\_pauling\_test}, \textit{include\_alloys}, \textit{check\_metallicity}, \textit{oxidation\_states\_set}, \textit{consensus}, \textit{include\_zero}, \textit{commonality}
\If{$c$ has only one element} \State \textbf{return} True \EndIf
\If{\textit{include\_alloys} \textbf{and} all elements in $c$ are metals} \State \textbf{return} True \EndIf
\If{\textit{check\_metallicity} \textbf{and} $\mathrm{metallicity}(c) \ge \mathrm{threshold}$} \State \textbf{return} True \EndIf
\State $(elements, stoich) \gets \mathrm{reduced\_stoichiometry}(c)$
\If{\textit{oxidation\_states\_set} is None}
    \State $\Omega_{\text{filtered}} \gets \mathrm{ICSD24Filter}(consensus, include\_zero, commonality)$
\Else
    \State $\Omega_{\text{filtered}} \gets \mathrm{preset\_ox\_states}$
\EndIf
\For{each $(q_1,\dots,q_n) \in \prod_i \Omega_{\text{filtered}}(E_i)$}
    \If{$\lnot\, \mathrm{charge\_neutral}(\{q_i\}, stoich)$}
        \State \textbf{continue}
    \EndIf
    \If{\textit{use\_pauling\_test}}
        \State $en\_ok \gets \mathrm{pauling\_test}(\{q_i\}, elements)$ \Comment{fallback to True on missing EN}
        \If{$\lnot\, en\_ok$}
            \State \textbf{continue}
        \EndIf
    \EndIf
    \State \textbf{return} True \Comment{short-circuit}
\EndFor
\State \textbf{return} False
\end{algorithmic}
\end{algorithm}

\subsection{Benchmarking generative models}

Validity rates in Tables~\ref{table_nonunique} and~\ref{table_unique} were computed with \texttt{SMACT} v4.0.0 \cite{davies_computational_2016}, setting \texttt{include\_alloys=True}, \texttt{check\_metallicity=True} and \texttt{include\_zero=False}. The consensus threshold was fixed at $C=3$ for the oxidation-state columns and the commonality threshold $\theta$ varied as reported; the $q$ and $q+\chi$ columns use $C=0$ with the Pauling electronegativity rule disabled and enabled respectively. For each model, 10,000 compositions were sampled without conditioning. For the unique-composition analysis (Table 2), samples were deduplicated by reduced formula and 7,000 unique compositions were then drawn at random from each model's deduplicated set, so that all models are compared on an equal sample size.

\subsection{Thermodynamic stability evaluation}
Generated structures were relaxed with the MACE-MPA-0 (medium) foundation interatomic potential\cite{batatiaFoundationModelAtomistic2025} in float64 precision, using batched FIRE optimisation with a Frechet cell filter, which relaxes atomic positions and lattice parameters simultaneously, as implemented in torch-sim\cite{cohenTorchSimEfficientAtomistic2025}. The energy above the convex hull was computed for each relaxed structure using pymatgen's PatchedPhaseDiagram, constructed from all Materials Project entries (snapshot of 9 April 2025, archived at \texttt{doi.org/10.6084/m9.figshare.30589436}) with uncorrected GGA/GGA+U total energies, consistent with the uncorrected DFT energies on which MACE-MPA-0 is trained; this follows the convention of Matbench Discovery\cite{riebesellFrameworkEvaluateMachine2025}. Values below the current hull were retained rather than clipped at zero. Structures whose MACE relaxation failed to converge (25 structures, 0.04\%) and all Yb-containing compositions (1,506 structures, 2.5\%), for which the Materials Project snapshot contains no reference entries and E$_{hull}$ is therefore undefined, were excluded, leaving n = 58,469 structures for the analysis in Figure~\ref{ehulls}. A structure is considered metastable when E$_{hull}$ < 0.1 eV/atom. Because MLIP-predicted formation energies carry errors of a few tens of meV/atom, individual E$_{hull}$ values are approximate; the analysis in Figure~\ref{ehulls} therefore rests on distributional comparisons between valid and invalid subsets rather than per-compound stability claims.

\subsection{Reinforcement-learning fine-tuning}

The pretrained Chemeleon2 latent diffusion model \cite{parkGuidingGenerativeModels2025a} was fine-tuned under three reward formulations. The binary reward is the validity operator itself, $R_{\mathrm{bin}}(c)=\mathbf{V}(c)$, with $\Omega_{\text{filtered}}(E_i)$ constructed from the \texttt{icsd24} oxidation-state set. The continuous reward $R_{\mathrm{cont}}(c)$ scores a composition by how strongly its charge-neutral assignments rely on commonly observed species, using the per-element ICSD24 proportions $f(E_i^{q_i})$ (the fraction of element $E_i$'s recorded occurrences in oxidation state $q_i$; $C=3$, \texttt{include\_zero=False}),
\begin{equation}
R_{\mathrm{cont}}(c)=
\begin{cases}
1, & \text{single element, or all elements metallic,}\\[4pt]
0, & \exists\, E_i \notin \text{ICSD24 table,}\\[4pt]
\displaystyle\max_{(q_1,\dots,q_n)\in \mathcal{N}(c)}
\left(\prod_{i=1}^{n}\frac{f(E_i^{q_i})}{100}\right)^{1/n}, & \text{otherwise,}
\end{cases}
\label{eq:cont_reward}
\end{equation}
where $n$ is the number of distinct elements and $\mathcal{N}(c)$ is the set of charge-neutral assignments, each $q_i$ restricted to that element's recorded states.

The evaluation-metric reward is the default Chemeleon2 ``DNG'' objective \cite{parkGuidingGenerativeModels2025a} with a weighted sum of a creativity term (uniqueness and novelty via the minimum AMD distance to same-formula references), an energy term (clamped, negated $E_{hull}$), and structural and compositional diversity terms (maximum mean discrepancy against the reference set), with weights $1.0$, $1.0$, $0.1$ and $1.0$. A structure is metastable when $E_{hull}<0.1$~eV/atom against a Materials Project phase diagram, unique when it is not a duplicate of another generated structure, and novel when it does not appear in the MP-20 reference set; mSUN is the fraction of structures that are simultaneously metastable, unique and novel.

All three rewards were optimised with Group Relative Policy Optimisation (GRPO) \cite{shaoDeepSeekMathPushingLimits2024} using a clipped surrogate objective, group-relative advantage normalisation over groups of 64 samples, a $k_3$-estimator KL penalty to the pretrained model (weight $1.0$) and an entropy bonus (weight $10^{-5}$). Generations used DDIM \cite{songDenoisingDiffusionImplicit2020a} sampling with 50 steps. Optimisation used AdamW \cite{loshchilovDecoupledWeightDecay2019} with learning rate $10^{-5}$, no weight decay, a constant schedule and gradient clipping at norm $1.0$, batch size 5 and 400 update steps.
\section*{Data availability}
The data and scripts required to reproduce the figures and analyses in this study are available at \url{https://github.com/KingaMas/ultra-fine-filters}.

\section*{Code availability}
The \texttt{SMACT} library used for filtering, analysis, and materials screening is openly available at \url{https://github.com/WMD-group/SMACT}.

\section*{Acknowledgements}
This work was supported by EPSRC project EP/X037754/1. The authors also benefited from the AI for Chemistry: AIchemy hub (EPSRC grant EP/Y028775/1 and EP/Y028759/1). The authors thank Junkil Park for valuable feedback and suggestions that improved the manuscript.

\section*{Author contributions}
K.O.M.: Conceptualisation, methodology, formal analysis, software, visualisation, writing -- original draft. P.D.: Formal analysis, visualisation, writing -- review \& editing. H.P.: Supervision. A.O.: Software. M.N.: Data curation, writing -- review \& editing. A.W.: Conceptualisation, supervision, funding acquisition, writing -- review \& editing.

\section*{Competing interests}
A.W. is Chief Scientific Officer at CuspAI. A.O. is a Machine Learning Scientist at MatNex. 
The remaining authors declare no competing interests.

\bibliographystyle{naturemag-npj-3}
\bibliography{references}

\clearpage
\setcounter{figure}{0}
\renewcommand{\thefigure}{S\arabic{figure}}
\section*{Supplementary Information}
\renewcommand{\thetable}{S\arabic{table}}
\setcounter{table}{0}   

\begin{figure}[!htbp]
    \centering
    \includegraphics[width=\linewidth]{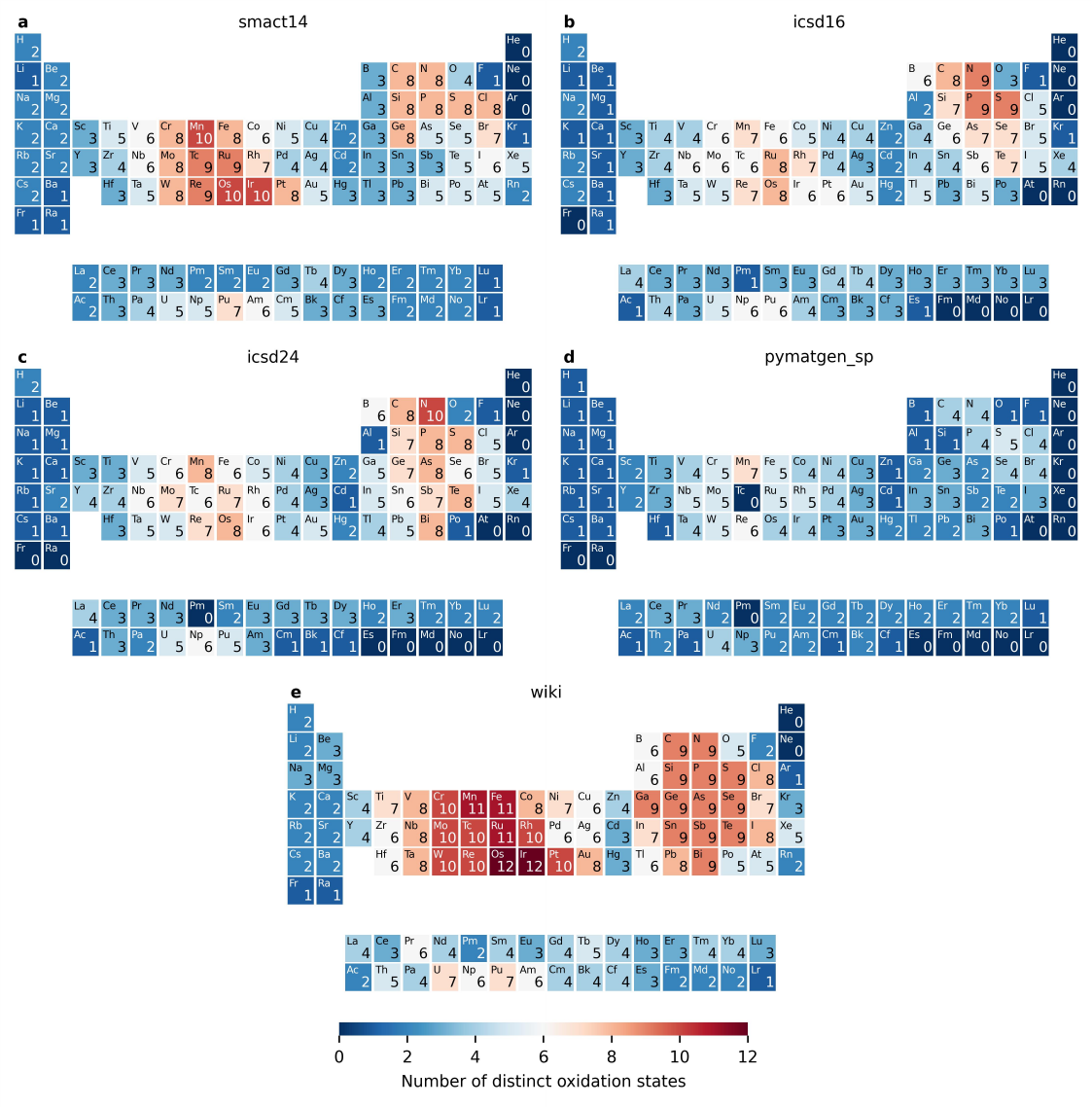}
    \caption{Number of oxidation states listed per element in each of the five built-in oxidation-state lists. Each list is a fixed catalogue compiled from a different source, and no further filtering is applied: (a) the original SMACT list, (b) a list drawn from the ICSD in 2016, (c) an updated ICSD list from 2024, (d) the list used by the pymatgen structure predictor, and (e) the list reported on Wikipedia. The colour and the number in each tile show how many oxidation states are listed for that element, with navy tiles marking elements that are absent from the corresponding list.}
    \label{fig:element_frequency2}
\end{figure}

\subsection*{Computational cost of chemical screening}

\begin{table}[!htbp]
\centering
\caption{Single-core timing benchmark for \texttt{SMACT} screening of generated compositions. Timings were measured using the same six 10{,}000-composition model outputs and the same screening path used for Table~\ref{table_nonunique}. Each model and filter setting was screened sequentially in a single Python process with \texttt{OMP\_NUM\_THREADS=1}, after a warm-up pass, and each timing was repeated five times. Values report the mean wall time across the six model sets. Across all model and filter combinations, the full observed range was 121--262~s per 10{,}000 compositions, corresponding to 12.1--26.2~ms per composition and 38--82 compositions s\(^{-1}\).}
\label{tab:screening_cost}
\small
\begin{tabular}{lccc}
\toprule
Filter setting & 
\shortstack{Mean wall time for\\10{,}000 compositions / s} &
\shortstack{Mean time per\\composition / ms} &
\shortstack{Mean throughput\\/ compositions s\(^{-1}\)} \\
\midrule
\(q\) & 155 & 15.5 & 64.5 \\
\(q+\chi\) & 183 & 18.3 & 54.6 \\
\(C=3\) & 166 & 16.6 & 60.2 \\
\(C=3,\theta=0.05\) & 197 & 19.7 & 50.8 \\
\(C=3,\theta=0.10\) & 202 & 20.2 & 49.5 \\
\(C=3,\theta=0.50\) & 148 & 14.8 & 67.6 \\
\bottomrule
\end{tabular}
\end{table}

The benchmark was run on an AMD EPYC 7742 64-core processor using Python~3.12.12, \texttt{SMACT}~4.0.0 and \texttt{pymatgen}~2025.10.7. One CPU core was used per timed screening process. The reported timings are conservative for the current implementation, because the manuscript screening path reconstructs the ICSD24 oxidation-state table on each call rather than caching the filtered table between compositions.

\end{document}